\documentclass[sigconf,screen,nonacm]{acmart}

\setcopyright{rightsretained}
\acmConference[NordiCHI '26]{Proceedings of the Nordic Conference on Human-Computer Interaction}{October 5--7, 2026}{Vaasa, Finland}
\acmBooktitle{Proceedings of the Nordic Conference on Human-Computer Interaction (NordiCHI'26), October 5--7, 2026, Vaasa, Finland}
\copyrightyear{2026} 
\acmYear{2026}

\graphicspath{{figs/}{figures/}{pictures/}{images/}{./}} 

\usepackage{booktabs}
\usepackage{soul} 
\usepackage{subcaption} 
\usepackage{tikz}
\usepackage{tcolorbox}
\tcbuselibrary{skins}

\usepackage{cleveref}
\crefname{figure}{Figure}{Figures}
\Crefname{figure}{Figure}{Figures}
\crefname{section}{Section}{Sections}
\usepackage{hyperref}

\hypersetup{colorlinks=true}

\newif\ifrevisions
\revisionstrue 

\ifrevisions

   \newcommand{\del}[1]{\textcolor{red!75!black}{\st{#1}}}
\else

  \newcommand{\del}[1]{} 
\fi

\NewTotalTColorBox[auto counter]{\Definition}{ +m }{ 
    notitle,
    colback=orange!5!white,
    frame hidden,
    boxrule=0pt,
    enhanced,
    sharp corners,
    borderline west={4pt}{0pt}{orange!50!blue},
}{
    \textcolor{orange!50!blue}{
        \sffamily
        \textbf{Definitions.}
    }%
    #1
}

\usepackage{fontawesome5}
\usepackage{xcolor}
\usepackage{etoolbox}

\definecolor{channelInput}{HTML}{2E5C8A}
\definecolor{channelOutput}{HTML}{C9803A}

\newcommand{\channelIcon}[1]{%
  \ifstrequal{#1}{pointing}{\faMousePointer}{%
  \ifstrequal{#1}{touch}{\faHandPointer}{%
  \ifstrequal{#1}{voice}{\faMicrophone}{%
  \ifstrequal{#1}{gaze}{\faEye}{%
  \ifstrequal{#1}{gesture}{\faHandPaper}{%
  \ifstrequal{#1}{body}{\faWalking}{%
  \ifstrequal{#1}{tangible}{\faCube}{%
  \ifstrequal{#1}{keyboard}{\faKeyboard}{%
  \ifstrequal{#1}{visual}{\faDesktop}{%
  \ifstrequal{#1}{audio}{\faVolumeUp}{%
  \ifstrequal{#1}{haptic}{\faMobile}{%
  \ifstrequal{#1}{olfactory}{\faSpa}{%
  \textbf{?#1?}}}}}}}}}}}}}%
}

\newcommand{\channelLabel}[1]{%
  \ifstrequal{#1}{pointing}{Pointing}{%
  \ifstrequal{#1}{touch}{Touch}{%
  \ifstrequal{#1}{voice}{Voice}{%
  \ifstrequal{#1}{gaze}{Gaze}{%
  \ifstrequal{#1}{gesture}{Gesture}{%
  \ifstrequal{#1}{body}{Body}{%
  \ifstrequal{#1}{tangible}{Tangible}{%
  \ifstrequal{#1}{keyboard}{Keyboard}{%
  \ifstrequal{#1}{visual}{Visual}{%
  \ifstrequal{#1}{audio}{Audio}{%
  \ifstrequal{#1}{haptic}{Haptic}{%
  \ifstrequal{#1}{olfactory}{Olfactory}{%
  #1}}}}}}}}}}}}%
}

\newcommand{\ic}[1]{{\color{channelInput}\channelIcon{#1}}}
\newcommand{\oc}[1]{{\color{channelOutput}\channelIcon{#1}}}
\newcommand{\icL}[1]{{\color{channelInput}\channelIcon{#1}\,\channelLabel{#1}}}
\newcommand{\ocL}[1]{{\color{channelOutput}\channelIcon{#1}\,\channelLabel{#1}}}

\begin{document}


\title{Channels and Substrates: Distributed Cognition as an Interaction Model for Ubiquitous Analytics}

\author{Niklas Elmqvist}
\email{elm@cs.au.dk}
\orcid{0000-0001-5805-5301}
\affiliation{
  \institution{Aarhus University}
  \city{Aarhus}
  \country{Denmark}
}

\author{Panagiotis D.\ Ritsos}
\email{p.ritsos@bangor.ac.uk}
\orcid{0000-0001-9308-3885}
\affiliation{
  \institution{Bangor University}
  \city{Bangor}
  \country{United Kingdom}
}

\author{Peter W.\ S.\ Butcher}
\email{p.butcher@bangor.ac.uk}
\orcid{0000-0002-3361-627X}
\affiliation{
  \institution{Bangor University}
  \city{Bangor}
  \country{United Kingdom}
}

\begin{abstract}
    Traditional HCI interaction models assume a single monolithic interface and a stable sensorimotor loop. 
    These models fit poorly with cross-device (XVA) and ubiquitous analytics (UA), where interactive data sensemaking unfolds across multiple devices, artifacts, and people in disparate settings from the office to the factory floor.
    In this paper, we show how interaction in ubiquitous analytics can be modeled using distributed cognition as propagation of representational state across substrates---minds, speech, bodies, artifacts, and devices---rather than as traffic through a single interface.
    On this basis we introduce input and output channels as generalizations of the visual channels from data visualization: just as visual channels carry data through properties of the visual substrate, input and output channels carry representational state through substrates whose availability, suitability, and preferability depend on context.
    We demonstrate the channels and substrates framework by reanalyzing several ubiquitous, immersive, and situated analytics systems. 
\end{abstract}



\begin{teaserfigure}
    \centering
    \includegraphics[width=\linewidth, alt={Teaser.}]{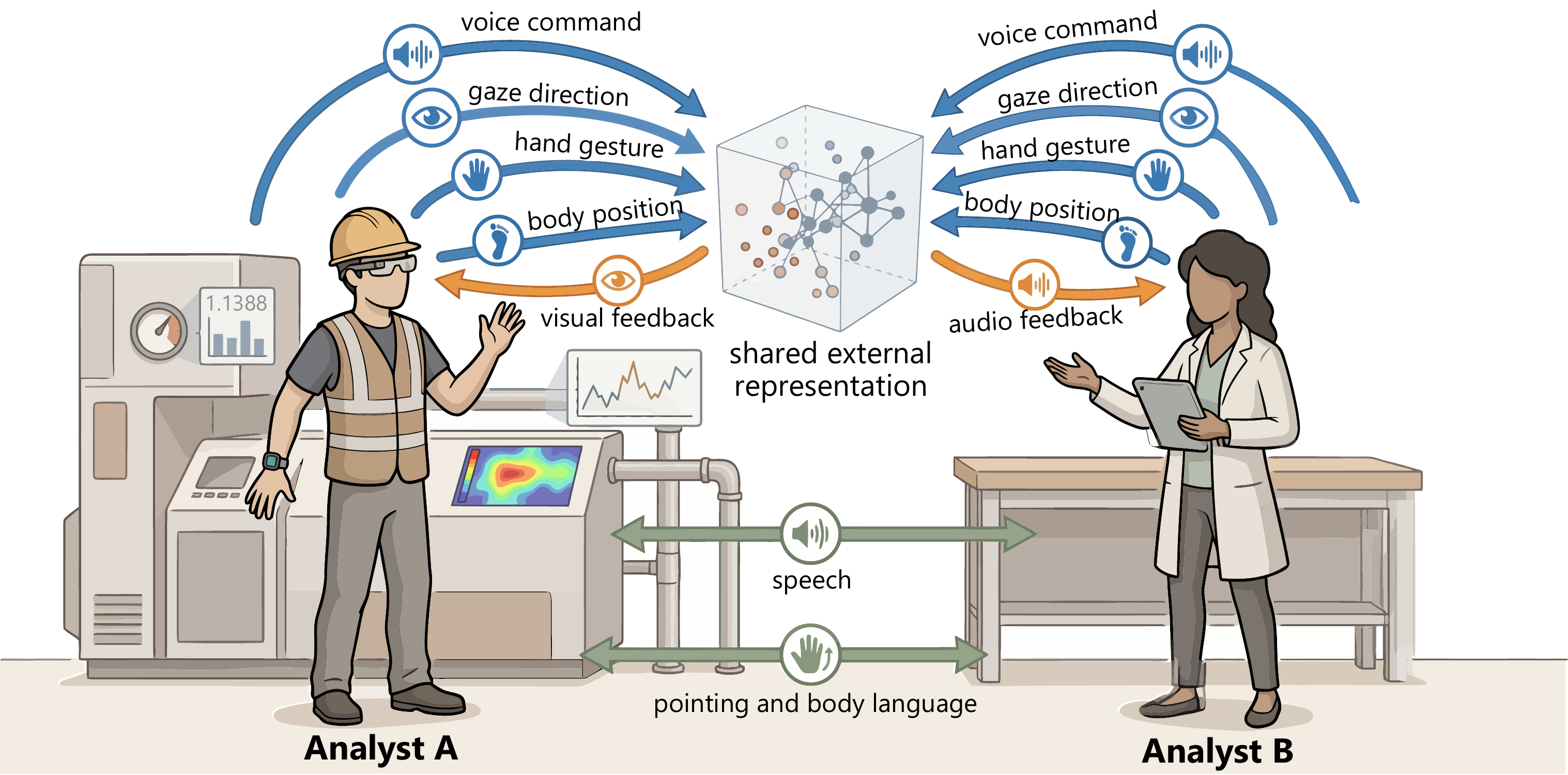}
    \caption{%
        \textbf{Interaction as distributed cognition across input channels and substrates.}
        Two analysts collaborate around a shared external representation (a 3D node-link visualization, center).
        Each analyst propagates representational state to the system through multiple input channels---voice command, gaze direction, hand gestures, and body position (blue arrows)---and receives output via visual and audio feedback channels (orange arrows).
        The analysts also communicate directly with each other through speech and through pointing and body language (green arrows).
        The shared visualization, the ambient displays in the environment (heatmap, line chart), the analysts' bodies, and the physical artifacts they hold together constitute the distributed cognitive system in which sensemaking unfolds.
    }
    \Description{Illustration of two analysts collaborating in an industrial setting around a shared 3D scatterplot with node-link connections shown at the top center. Analyst A, on the left, wears a hard hat, safety glasses, smart glasses, a high-visibility vest, and a smartwatch, and stands next to industrial equipment that displays a gauge, a bar chart, and a heatmap. Analyst B, on the right, wears a lab coat and holds a tablet, standing beside a workbench; a line-chart display is mounted between them. Four blue arrows labeled voice command, gaze direction, hand gesture, and body position flow from each analyst up to the shared 3D visualization in the center. Two orange arrows labeled visual feedback and audio feedback flow from the visualization back to the analysts. Two green bidirectional arrows between the analysts are labeled speech and pointing and body language. Each channel is marked with a small circular icon: a speaker for voice and audio, an eye for gaze and visual, a hand for gesture, and a footprint for body position.}
    \label{fig:teaser}
\end{teaserfigure}

\maketitle

\section{Introduction}

Interaction models in HCI have historically assumed a single device, a single user, and a stable sensorimotor loop. 
Norman's seven-stage action cycle~\cite{Norman1988}, for example, describes a closed feedback circuit between a person and an artifact: form a goal, execute an action, evaluate the result.
This framing has served the field well for decades~\cite{Shneiderman2016dtui}.
Hornbæk and Oulasvirta~\cite{DBLP:conf/chi/HornbaekO17} sharpen the question of what interaction is, identifying multiple competing accounts (dialogue, tool use, optimal behavior, embodiment) without converging on one. 
Bergström and Hornbæk~\cite{DBLP:journals/ijmms/BergstromH25} propose the DIRA model to formalize the user interface itself as an assembly of devices, interaction techniques, and representations.
Both efforts expose a shared gap: mainstream models describe the \textit{components} and \textit{structure} of interaction but say little about what \textit{flows} between human and system, through which pathways, and why some pathways work in one context but collapse in another.

Ubiquitous analytics (UA)---data sensemaking that unfolds across smartphones, tablets, wall displays, AR headsets, and tangible objects in mobile, immersive, and situated contexts~\cite{Elmqvist2013, Elmqvist2023}---is precisely such a setting.
An emergency manager at a wall display, a marine biologist analyzing reef data through AR on a rocking boat, and a factory technician querying machine performance while her hands hold tools all perform analytical interaction, yet no single-device model captures the shifting assemblies of bodies, artifacts, and displays they work with.

Data visualization has already identifies the graphical attributes that carry information from data to the human visual system---position, length, color, texture, size---and calls these \textit{visual channels}~\cite{Munzner2014, Bertin1967}.
A bar chart encodes a quantitative value in the height of a rectangle; that height is a ``channel'' running over the substrate of light between screen and retina.
The practice is so ingrained that no visualization paper describes an encoding without naming which channel carries which variable.
However, even if the translation seems obvious, there is no equivalent on the input side.
HCI catalogs input modalities (gesture, gaze, voice, touch) and input devices (controller, keyboard, hand tracker), and DIRA~\cite{DBLP:journals/ijmms/BergstromH25} organizes these into assemblies, but neither tradition names the structured information-carrying pathways that connect analyst intent to system state.
In this paper, we generalize the visual-channel concept into \textit{input channels} and \textit{output channels}: structured pathways for propagating representational state between human and computational substrates.
We ground this generalization in \textit{socially distributed cognition}~\cite{Hutchins1995, Liu2008, DBLP:journals/tochi/HollanHK00}, which models cognition as the propagation of representations across media.
From this perspective, \textit{interaction} is information transfer across \textit{representational substrates}~\cite{Mackay2025substrates}, and \textit{interfaces} are the boundaries where that transfer transforms one representation into another.
Channels are, therefore, the specific attributes of a substrate that carry information across an interface, just as height is the attribute of a bar that carries a quantitative value across the visual interface between screen and eye.

We operationalize this framework as follows (see Figure~\ref{fig:teaser}).
A channel is defined by its \textit{substrate}~\cite{Mackay2025substrates}: the physical medium it runs over (air for gesture, light for gaze, acoustic energy for voice, contact for touch).
Each channel has several intrinsic properties:
\textit{bandwidth} (information carried per unit time), \textit{precision} (achievable granularity), and \textit{directness} (how closely action in the channel maps onto the representation space, following Beaudouin-Lafon's~\cite{Beaudouin-Lafon2000} degree of indirection). 
Finally, each channel has a \textit{context-dependent status}---available, suitable, or preferable---determined by the physical, social, and task setting the analyst occupies.
A gestural channel runs over the motor-proprioceptive substrate, offers high directness and moderate bandwidth, but its status shifts: available on a factory floor, unsuitable when the worker's hands hold tools, preferable at a wall display where voice would be socially disruptive.
To demonstrate that this frame does analytical work beyond taxonomy, we reanalyze several published systems---from Bolt's ``Put That There''~\cite{Bolt1980} to Wizualization~\cite{DBLP:journals/tvcg/BatchBRE24}---and show that the channel-substrate vocabulary surfaces design rationale that the original papers left implicit or described ad hoc.

The contributions of this paper include the following:
(1) a theoretical framework grounding interaction in distributed cognition and generalizing visual channels into input and output channels analyzed over substrates;
(2) an operationalization of input and output channels and their contextual availability, suitability, and preferability; and
(3) a reanalysis of several published systems demonstrating the framework's explanatory and generative capacity for ubiquitous, immersive, and situated analytics.

The remainder of this paper is structured as follows.
Section~\ref{sec:background} reviews the related work on interaction models, distributed cognition in HCI and visualization, and existing taxonomies of input modalities.
Section~\ref{sec:model} presents the channel-substrate framework and the available–suitable–preferable operationalization.
This is followed by Section~\ref{sec:channels}, which catalog some of the most common input and output channels in HCI, data visualization, and ubiquitous analytics.
Section~\ref{sec:analysis} applies the framework to existing systems, showing the analytical work it performs. 
Section~\ref{sec:discussion} discusses implications for UA interaction design and limitations of the framework. Section~\ref{sec:conclusion} concludes.
\section{Background}
\label{sec:background}

Here we review the foundations our model builds on: interaction models from HCI, interaction in data visualization, the ubiquitous analytics research area, and distributed cognition.

\subsection{HCI Interaction Models}
 
Norman's seven-stage action cycle~\cite{Norman1986, Norman1988} remains the dominant account of how people interact with systems.
The model decomposes interaction into a loop: form a goal, plan the action, specify it, execute it, perceive the result, interpret it, evaluate it against the goal.
Two ``gulfs'' separate intention from outcome: the \textit{gulf of execution} (translating intent into action) and the \textit{gulf of evaluation} (interpreting feedback).
Good HCI design narrows both gulfs.
The model has shaped decades of interface research and still anchors textbook treatments of HCI~\cite{Shneiderman2016dtui}.
 
Shneiderman's direct manipulation~\cite{DBLP:journals/computer/Shneiderman83,Shneiderman2016dtui} operationalized Norman's ideas for graphical interfaces: continuous visual representation of objects, rapid and reversible physical actions, and incremental feedback.
The paradigm works well for desktop systems, where a single display, a mouse, and a keyboard form a stable sensorimotor loop.
 
Beaudouin-Lafon's instrumental interaction~\cite{Beaudouin-Lafon2000,Beaudouin-Lafon2004} moved beyond direct manipulation by separating the \emph{instrument}---a mediating object that transforms user input into commands on domain objects---from the objects it acts upon.
Instruments can vary in degree of indirection and degree of integration, providing an analytic vocabulary for comparing interaction designs.
More recently, Mackay and Beaudouin-Lafon~\cite{Mackay2025substrates} extended the model through \emph{interaction substrates}: computational media that hold digital information, apply constraints and transformations to it, and generate information consumable by other substrates.
Our use of the term \emph{substrate} follows theirs, but we connect it to distributed cognition.
 
Hornb{\ae}k and Oulasvirta~\cite{DBLP:conf/chi/HornbaekO17} cataloged seven competing accounts of what ``interaction'' means in HCI---from pragmatic dialogue to phenomenological engagement to information processing---and showed that the community lacks consensus on what exactly it studies.
Bergstr\"om and Hornb{\ae}k's DIRA model~\cite{DBLP:journals/ijmms/BergstromH25} addressed this by formalizing the user interface as a dynamic assembly of representational components.
DIRA is a parallel effort to our model; our contribution is to ground the assembly in distributed cognition and to operationalize it for ubiquitous analytics through a channel vocabulary.
 
These models share a common assumption: a single user, a single device, a stable context.
The action cycle loops through one interface and one user.
Instrumental interaction mediates between one user and one application.
Even DIRA, despite modeling the interface as an assembly, presupposes a bounded system.
This assumption breaks when analysis moves off the desktop.

\subsection{Interaction for Visualization}

Interaction has historically received less attention than visual encoding in the visualization community.
Yi et al.~\cite{Yi2007} documented this imbalance through a review of 59 papers and 51 systems, distilling seven categories of interaction intent: Select, Explore, Reconfigure, Encode, Abstract/Elaborate, Filter, and Connect.
Their taxonomy describes \emph{what} users want to do, but not \emph{how} those intents map onto the physical and cognitive resources available in a given setting.
 
Pike et al.~\cite{Pike2009} argued for a ``science of interaction'' in visual analytics, framing interaction as the mechanism through which knowledge is constructed, tested, and shared.
They called for interaction research that moves beyond event-level descriptions toward cognitively grounded principles.
Dimara and Perin~\cite{DBLP:journals/tvcg/DimaraP20} tackled the definitional question head-on, synthesizing views from the visualization and HCI communities into a definition of interaction as intentional dialogue that changes what is seen and understood.
Brehmer and Munzner~\cite{Brehmer2013} contributed a multi-level task typology distinguishing \emph{why} a task is performed from \emph{how} and \emph{what}, filling the gap between low-level interaction events and high-level analytical goals.
 
Yalcin et al.~\cite{DBLP:conf/beliv/YalcinEB16} applied Norman's action cycle to visual data exploration, mapping the seven stages onto cognitive stages in the analytical process.
Lam~\cite{DBLP:journals/tvcg/Lam08} extended Norman's two gulfs to three by adding the gulf of goal formation, which is particularly relevant for analytical work where knowing \emph{what to look for} is often harder than knowing how to look.
 
These contributions map the terrain of visualization interaction.
Yet they share a common limitation: they assume a desktop analyst with mouse and keyboard.
We have taxonomies of intents and task types, but no systematic way to describe the properties of the channels through which analytical intent reaches the system, or through which the system's response reaches the analyst.
Our model addresses this gap. 

\subsection{Ubiquitous, Immersive, and Situated Analytics}

Ubiquitous analytics (UA)~\cite{Elmqvist2013} is the overarching research area of data visualization and analytics that leverages post-WIMP devices and displays, such as mobile devices, large-display environments, and XR technologies.
Elmqvist and Irani defined the area and its scope in 2013; Elmqvist~\cite{Elmqvist2023} later reviewed its full ambition in 2023, arguing that data analytics should be available anywhere and everywhere, freed from the desktop workstation that still dominates practice.
 
Two subfields build on this framework.
Immersive analytics (IA)~\cite{Chandler2015, Marriott2018} focuses specifically on mixed, augmented, and virtual reality as platforms for data exploration.
Situated analytics (SA)~\cite{elsayed16situateddef, Shin2024} anchors analysis in the physical space to which the data refers: sensor readings examined at the sensor, production metrics reviewed on the factory floor.
Both inherit UA's departure from desktop assumptions: the analyst may be standing, walking, or riding a boat; the display may be a headset, a handheld screen, or a wall; input may arrive through gesture, gaze, voice, or body position.
The environment imposes constraints---noise, vibration, social norms, safety requirements---that modulate which interaction channels are usable at any given moment.
 
Existing work on interaction in these settings tends toward either system descriptions or modality-specific surveys.
What is missing is a unifying interaction model that explains \emph{why} certain channels work in certain settings, \emph{why} they fail in others, and \emph{how} designers should reason about these tradeoffs.

\subsection{Distributed and 4E Cognition}

We draw on distributed cognition (DCog), introduced by Hutchins~\cite{Hutchins1995}, as the foundation for our model, and for ubiquitous analytics as a whole~\cite{Elmqvist2023}.
DCog models cognition as information flow across representational media---people, artifacts, instruments, environmental structures---distributed among different agents and embedded within sociotechnical systems.
Hutchins's paradigmatic example, the navigation team of a U.S.\ Navy ship, demonstrated how navigation emerges from coordinated transformations of representational state across charts, instruments, verbal reports, and mental computations; no single crew member possesses all the knowledge needed.
Hollan, Hutchins, and Kirsh~\cite{DBLP:journals/tochi/HollanHK00} brought DCog into HCI, arguing that the proper unit of analysis for interaction design is the entire functional system---user, tools, task environment---not the individual user.
 
Liu, Nersessian, and Stasko~\cite{Liu2008} applied DCog to data visualization, demonstrating that effective visualization emerges from the cognitive system formed by analyst, software, infrastructure, notebooks, and collaborators.
Information flows through this system---from notepad to verbal instruction, from spoken word to typed query, from query result to visual display, and back to handwritten notes---with each transformation preserving meaning while adapting to a different representational medium.
This transformation is not merely an aid to perception, memory, or computation---although Kirsh also argue for these exact benefits in our use of physical space~\cite{DBLP:journals/ai/Kirsh95}---but also a way to benefit from different situated cognitive strategies inherent to each representation~\cite{Suchman1987}.
It is the very same phenonema that Scaife and Rogers call \textit{external cogniton}~\cite{Scaife1996}, and which are often quoted as a main benefit of data visualization.
 
DCog belongs to a broader movement in cognitive science sometimes called 4E cognition: the view that cognition is \ul{e}mbodied~\cite{Shapiro2011}, \ul{e}mbedded, \ul{e}nactive, and \ul{e}xtended.
Clark and Chalmers's extended mind thesis~\cite{Clark1998} and Suchman's situated action~\cite{Suchman1987} contribute to this movement, as does Dourish's~\cite{Dourish2001} phenomenological account of embodied interaction, which makes action the basis for meaning rather than a downstream consequence of internal planning.
In the visualization community, embodied cognition has informed work on embodied human-data interaction~\cite{Elmqvist2011c} and embodied interaction instruments such as tangible lenses~\cite{Kim2012}.
Activity theory~\cite{Engestrom2015, Nardi1996} offers a complementary lens for tool-mediated work in organizational contexts.
B{\o}dker~\cite{DBLP:conf/nordichi/Bodker06} used it to analyze multi-artifact interaction in what she termed third-wave HCI.

\section{The Channels \& Substrates Model}
\label{sec:model}

The basic argument in this paper is to not privilege the human-computer interface over any other boundary at which information is transformed in the functional system that performs an analytical task.
A glance at a colleague, a note scribbled on paper, a reading off a physical gauge, and a drag-and-drop gesture on a tablet are all, from this vantage point, the same kind of event.
Drawing on concepts from data visualization, we argue in this section that distributed cognition can serve as a unified and comprehensive interaction model for human-computer interaction, and that this unification is particularly productive for the anytime, anywhere settings~\cite{Weiser1991} that motivate ubiquitous analytics~\cite{Elmqvist2013}.
Figure~\ref{fig:teaser} previews the view we develop: two analysts, several devices and physical artifacts, and a shared visualization, all coupled through multiple channels of information flow that our model treats on equal footing.

\subsection{Distributed Cognition as an Interaction Model}

As we have seen, distributed cognition~\cite{Hutchins1995, DBLP:journals/tochi/HollanHK00} holds that all information transfer between two representational media is an \textit{interaction}, and accordingly the boundary between two media is an \textit{interface}.
The media are said to be brought in \textit{coordination} with each other.
When a person writes a note on paper, their brain transforms symbolic meaning into movement (one interaction) of their fingers gripping the pen (another) which transfers strokes forming words to the page (a third).
Countless feedback interactions occur as the writer reads and confirms what they have written.
When that person uses a phone to write the same note, the process is similar even if the exact interactions and media differ: the fingers move not to draw strokes on paper using a pen, but to tap out letters on a keyboard.
Both are still interactions.

This means that a distributed cognition view of human-computer interaction does not privilege the interaction conducted through a user interface.
It also does not constrain interaction to a single entity or a single individual.
Based on this model, computational devices may have many interfaces of many forms, and the digital link between human and computer is only one among several representation changes in the full process.
Considering not just that single link, but the entire functional system---person, tools, physical environment---reveals interactions that conventional HCI interaction models overlook: the analyst reading a physical gauge, glancing at a colleague's gesture, and repositioning their body to change perspective.
It also means that there is no single monolithic interface, but instead a multitude of individual interfaces between the wide range of media involved in the interactive system, only some of them digital. 

In this paper, we draw on instrumental interaction~\cite{Beaudouin-Lafon2000, Beaudouin-Lafon2004} and more recent research on interaction substrates~\cite{Mackay2025substrates} to adopt the unified term \textit{substrate} to represent what distributed cognition calls a representational medium.
Substrates come in many shapes and forms; some are intrinsic to our physical world, such as the atmosphere surrounding us (which can convey soundwaves), the light we use to see (which can either reflect on diffuse surfaces or be directly generated by emissive ones), and the the air we breathe (which can convey chemical volatiles that carry smell).
Some are physical and human-made: a printed page, a road sign, a chalk mark on a factory floor, a hand-drawn diagram on a whiteboard, a paper flight progress strip annotated by an air traffic controller.
And others still are digital: the state of a variable in a running program, the pixels of a display, the registered touch points on a capacitive screen.
The latter are the focus of this paper.

\Definition{\textbf{Concepts from distributed cognition in HCI.} 
    \begin{itemize}
        \item \textit{Substrate}: a structured medium over which information can be represented and propagated.
        Substrates may be physical (air, light, contact, proprioception), artifactual (paper, signage, flight progress strips), or digital (memory, pixels, sensor readings).
        \item \textit{Interface}: the boundary between two substrates.
        \item \textit{Interaction}: the transformation and transfer of information across an interface.
    \end{itemize}
}

\subsection{What HCI Can Learn from Data Visualization}

The visualization community has long had a vocabulary for analyzing how information is carried over a specific substrate.
\textit{Visual channels}~\cite{Munzner2014}---also called ``visual variables''~\cite{Bertin1967}---are the attributes of geometric shapes through which data is encoded: position, length, angle, area, volume, color hue, color value, texture, shape.
The shapes themselves are called \textit{marks}, and early treatments of visualization use the term \textit{visual substrate} for the 2D or 3D space that contains them~\cite{Card1999}.
For example, a bar chart uses vertically oriented rectangles (marks) to convey quantitative values by their length (visual channel), measured from a common baseline.
Similarly, A pie chart uses circle segments (marks) arranged around a common center and conveys relative quantities through angular extent (visual channel).

A long and proud tradition of graphical perception experiments~\cite{Simkin1987} has worked out the perceptual ordering of these channels for different data types: nominal, ordinal, quantitative~\cite{Stevens1946}.
Bertin summarized centuries of cartography knowledge into an informal ranking~\cite{Bertin1967}.
Building on these findings, Cleveland and McGill in 1984 summarized results from a large number of graphical perception experiments to mostly confirm these guidelines~\cite{Cleveland1984}.
Mackinlay~\cite{Mackinlay1986} took the approach a step further by proposing an approach to automatic visualization based on the data types in a dataset.

Why does this matter for an interaction model?
Because, given an interaction model based on distributed cognition, we can see that a visual channel is basically just an attribute of a representational medium---or \textit{substrate} in the terminology of our model---that can convey meaning to a human observer.
For visualization, the substrate is a 2D or 3D display, the marks are geometric shapes arranged on it, and the channels are the mark attributes that encode data values.
Visual channels thus belong to geometric shapes (marks) displayed as a visual representation of a dataset; i.e., a visualization.
Generalized beyond the visual case, the same structure applies to any substrate: we can name the attributes of that substrate that carry information, and we can ask how much information each attribute carries and how well.
This is the move we make next.

\Definition{\textbf{Concepts from data visualization.} 
\begin{itemize}
    \item \textit{Data visualization}: a graphical, often interactive representation of data designed to aid cognition~\cite{Card1999, Munzner2014}.
    \item \textit{Visual mark}: a geometric shape representing one or more data points on a visual substrate.
    \item \textit{Visual channel}: an attribute of a visual mark---position, length, angle, area, color, shape, texture, orientation---that encodes information.
\end{itemize}
}

\subsection{Input and Output Channels}

Armed with knowledge from this detour into data visualization, we can now generalize from visual channels to any information-carrying substrate attribute.
An \textit{output channel} is an attribute of a substrate that carries information \textit{from} a computational device to a human user; visual channels are the best-studied special case.
An \textit{input channel} is an attribute of a substrate that carries information in the opposite direction, \textit{from} a human user \textit{to} a computational device.
Gesture, gaze, speech, touch, and body position are all input channels, each running over its own substrate.

Note that in the eyes of a distributed cognition model of human-computer interaction, there are many input and output channels involved in an interactive system.
Vocal chords generate sound waves over air; ears convert sound waves back into perceived words; the hand moves a pen over paper; the eye tracks a colleague's pointing finger.
Not every interactive system even contains a digital device.
Hutchins's analyses of ship navigation~\cite{Hutchins1995} and of airline cockpits~\cite{Hutchins1995cockpit} are interactive systems in exactly this sense.
So is air traffic control, where paper flight progress strips serve as the shared external substrate through which controllers coordinate aircraft movements~\cite{Mackay1999atc}.
And so, for that matter, is the exchange of orders, tickets, and payment in a busy restaurant kitchen.

Nevertheless, some of the substrates in a modern interactive system are indeed digital in nature, and these are the substrates whose channels we can engineer directly.
Traditional HCI calls the hardware attached to these substrates ``input devices'' and ``output devices.''
For this reason, a more precise name are \textit{digital input channels} and \textit{digital output channels}: attributes of a digital substrate (voltages on capacitive sensors, registered 3D positions, recognized phonemes, pixel color values) capable of carrying information to or from the computer.
These are the channels the system designer has direct control over, and for the remainder of this paper we simply use the terms \textit{input channel} and \textit{output channel} to refer specifically to the digital case.

\begin{figure*}[htb]
    \centering
    \includegraphics[width=\linewidth]{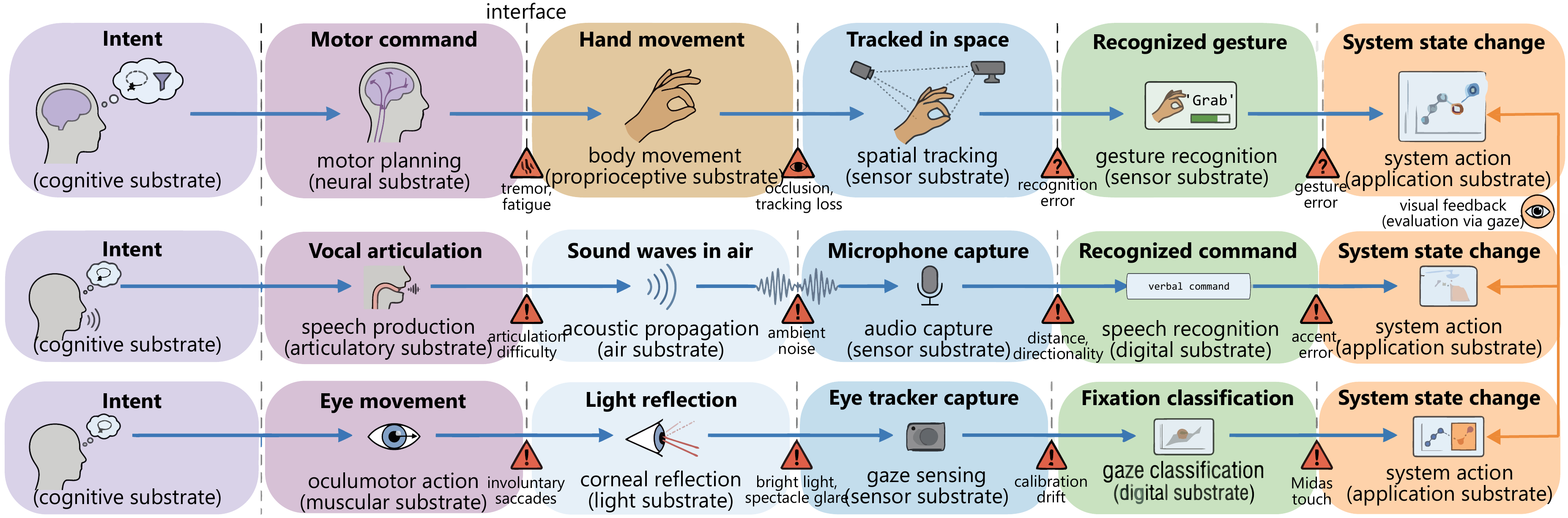}
    \caption{%
        \textbf{Input channels as chains of substrate transformations.}
        Three canonical input channels---gesture (top), voice (middle), and gaze (bottom)---are shown as parallel pathways propagating representational state from analyst intent to system action.
        Each channel passes through a sequence of substrates: cognitive (intent), neural or muscular (motor planning, oculomotor action, speech production), an environmental substrate carrying the signal (proprioception and air for gesture, light for gaze, acoustic waves for voice), a sensor substrate (spatial tracking, eye tracker, microphone), and a digital substrate where the signal is recognized and acted upon.
        Interfaces between substrates are the points where representational state transforms (and where it can be corrupted).
        Annotations below each interface name example failure modes at that boundary: tremor and fatigue in motor production, occlusion and tracking loss in sensing, ambient noise in acoustic propagation, Midas touch in gaze recognition, recognition error in the digital substrate.
        The figure illustrates that an input channel is not a single act but a chain, and that channel suitability depends on the reliability of every substrate in the chain.
    }
    \Description{Diagram showing three horizontal pathways arranged top to bottom: a gesture channel, a gaze channel, and a voice channel. Each pathway begins on the left with a box labeled Intent on a cognitive substrate and ends on the right with a box labeled system action on an application substrate. Between the endpoints, each channel passes through a sequence of labeled substrate boxes connected by short arrows. The gesture channel goes: Intent, motor planning on a neural substrate, body movement on a proprioceptive substrate, hand movement tracked in space on a sensor substrate, recognized gesture on a digital substrate, then system state change. The gaze channel goes: Intent, oculomotor action on a muscular substrate, eye movement, corneal reflection on a light substrate, eye tracker capture on a sensor substrate, fixation classification on a digital substrate, then system state change. The voice channel goes: Intent, speech production on an articulatory substrate, vocal articulation, sound waves in air on an acoustic substrate, microphone capture on a sensor substrate, recognized command on a digital substrate, then system state change. Below each transition between substrates, a small label names a failure mode: tremor and fatigue between motor planning and body movement; occlusion and tracking loss between body movement and spatial tracking; gesture error between tracking and recognition; articulation difficulty in speech production; ambient noise in acoustic propagation; accent and error in speech recognition; involuntary saccades in oculomotor action; bright light and spectacle glare in corneal reflection; calibration drift in eye tracker capture; Midas touch in fixation classification; recognition error and distance or directionality near the digital substrate. A verbal command example, the word Grab in quotes, is shown as an instance of recognized command. A visual feedback arrow labeled evaluation via gaze returns from the system to the analyst, indicating the evaluation side of the loop.}
    \label{fig:channel-overview}
\end{figure*}

Figure~\ref{fig:channel-overview} makes this concrete for three canonical input channels---gesture, gaze, and voice.
Each channel is a chain of substrate transformations that runs from analyst intent to a change in system state, and each interface along the chain is a point at which the representational state is transformed.
It is also a point where state can be transformed: tremor and fatigue at the motor interface; ambient noise in the acoustic substrate; calibration drift in eye tracking; Midas touch when fixations are misclassified as selections; ordinary recognition error at the final digital interface.
The view from distributed cognition is that an input channel is not a single act but an extended chain, and that the suitability of a channel for a given task depends on the reliability of every link in that chain.

\Definition{\textbf{Generalized state transfer.} 
\begin{itemize}
    \item \textit{Input channel}: an attribute of a digital substrate that carries information from a human user to a computational device (e.g., the registered 3D position of a tracked hand, the recognized command from a spoken utterance, the classified fixation from an eye tracker).
    \item \textit{Output channel}: an attribute of a digital substrate that carries information from a computational device to a human user (e.g., the color and position of a mark on a display, a haptic pulse on a wristband, a synthesized voice prompt).
\end{itemize}
}

\subsection{Channels in Ubiquitous Settings}
\label{sec:model:ubiquitous}

So far, you could argue that the introduction of these new concepts in this paper offer little more than new machinery with scant expressive power over existing traditional HCI models.
However, to this we would respond that a distributed cognition framing of human-computer interaction offers a more unified and comprehensive view of human interaction with a computer as a system of interacting components; some digital, some not.
The benefit is most clear in ubiquitous settings.
A conventional HCI model treats the digital output as a small porthole onto the computer's state and the digital input as a narrow return channel for user commands.
The distributed cognition view replaces the porthole with the whole functional system, including other actors, physical artifacts, and the history and culture of the practice in which the task is embedded~\cite{Hutchins1995}.
Norman's gulfs of execution and evaluation~\cite{Norman1988, Norman1986}, along with Lam's extension that adds the gulf of goal formation for analytical tasks~\cite{lam2008}, map naturally onto this expanded view: the execution gulf is closed by input channels, the evaluation gulf by output channels, and the goal formation gulf by the cognitive substrate in the analyst's brain and its complex interconnections with the external world as they make sense of the problem.

This expansion is particularly productive for ubiquitous, mobile, and cross-device settings, where multiple devices and multiple forms of interaction are the norm rather than the exception, and where digital output is integrated into the surrounding physical world rather than confined to a single screen.
These are also settings where there is an increasing need to integrate multiple individuals in collaboration.
A distributed cognition framing actively supports this integration by blurring the boundary between physical and digital worlds that traditional models presuppose~\cite{Elmqvist2023}.
In particular, it gives us the vocabulary to reason about which input and output channels are active in a given system, what attributes they carry information over, and, crucially, how the conditions of the time and place at which the system is used determine which channels can actually be used at all.

\begin{figure*}[htb]
    \centering
    \includegraphics[width=\linewidth]{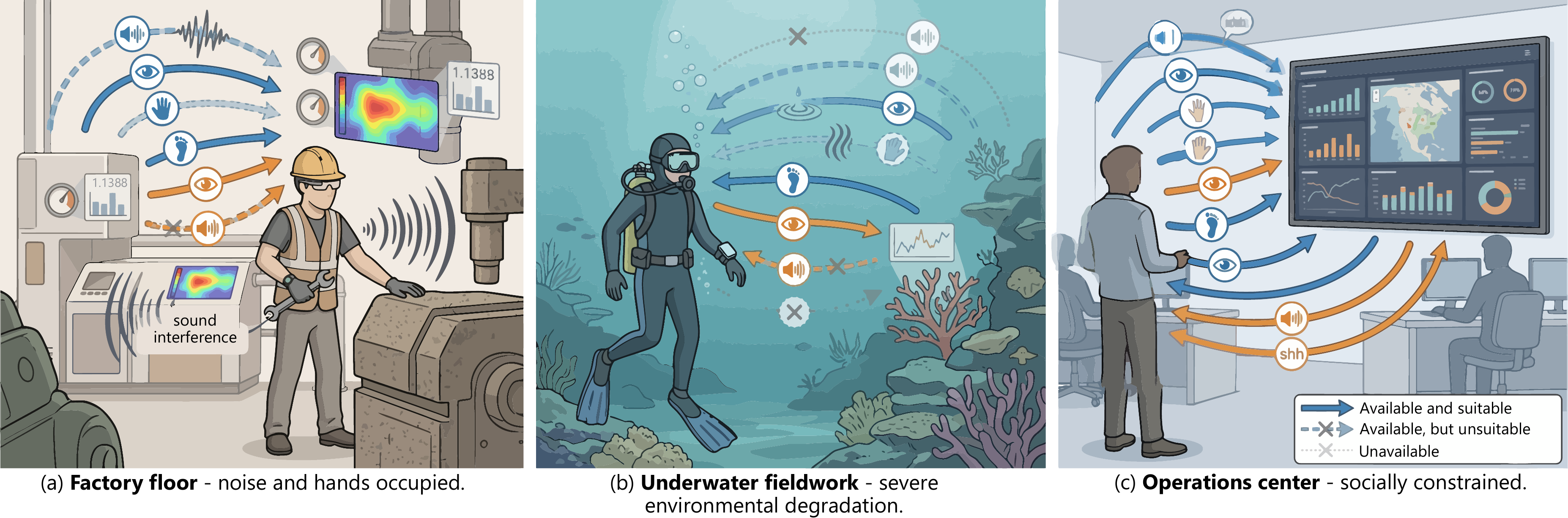}
    \caption{%
        \textbf{Channel availability and suitability shift across contexts.}
        Three analytical settings illustrate how the same input channels are reshaped by the physical and social substrates each context affords.
        (a) On the factory floor, the analyst's hands are occupied by tools and ambient machine noise degrades voice pickup; gaze and body position remain available and suitable, gesture and voice are available but unsuitable.
        (b) Underwater, water blocks acoustic propagation and restricts hand movement, the diving mask constrains gaze, and buoyancy limits locomotion; most channels are either unavailable or severely degraded.
        (c) In the operations center, every channel is technically available, but loud voice commands and large gestures violate the social norms of a shared professional space; subtle gaze and small body shifts remain suitable, while voice and broad gesture become unsuitable despite functioning.
        Green icons mark channels that are available and suitable, yellow icons mark channels available but unsuitable for the context, and red icons mark channels that are unavailable.
    }
    \Description{Figure with three panels arranged left to right, each showing an analyst in a different work context surrounded by small circular icons representing input channels. Each icon is color-coded: green for available and suitable, yellow for available but unsuitable, and red for unavailable. Panel a, labeled Factory floor with noise and hands occupied, shows a worker in a hard hat, safety glasses, smart glasses, and high-visibility vest standing next to industrial machinery that displays a numerical readout of 1.1388. Sound interference lines emanate from the machine. The worker's hands are holding tools. Icons around the worker show gaze and body position in green, and hand gesture and voice in yellow. Panel b, labeled Underwater fieldwork with severe environmental degradation, shows a diver in full scuba gear with mask, regulator, fins, and tank, surrounded by bubbles and fish. Icons around the diver show most channels in red, indicating unavailable, with only limited body position available. Panel c, labeled Operations center and socially constrained, shows an analyst in business attire standing in front of a large wall display of a world map with data overlays, with other colleagues visible at workstations in the background. Icons around the analyst show gaze and small body position in green as suitable, while voice command and large hand gesture icons are in yellow, indicating available but socially unsuitable. A legend at the bottom of the figure explains the three color categories: green for available and suitable, yellow for available but unsuitable, and red for unavailable.}
    \label{fig:settings}
\end{figure*}

Figure~\ref{fig:settings} shows why the last point matters.
The same three analytical settings, and roughly the same set of candidate input channels, yield three very different usable subsets.
On the factory floor (Figure~\ref{fig:settings}a), the analyst's hands are occupied by tools and ambient machine noise degrades voice pickup; gaze and body position remain usable.
Underwater (Figure~\ref{fig:settings}b), water blocks acoustic propagation, a diving mask constrains gaze, resistance and buoyancy limits locomotion, and thick gloves narrow the gesture vocabulary; most channels are degraded or unavailable.
In a shared operations center (Figure~\ref{fig:settings}c), every channel functions technically, but loud voice commands and large arm gestures violate the social norms of the space.
In all three cases, the relevant analytical distinction is not whether a channel exists but whether the substrate it runs over is actually there, whether the channel is appropriate to the task, and whether it is the best available choice given fatigue, social constraints, and the likelihood of interruption.
We develop these ideas next.

\section{Input and Output Channels}
\label{sec:channels}

Section~\ref{sec:model} defined a channel as an attribute of a substrate that carries information across an interface.
To put this definition to work we need two things: (1) a way to characterize channels that applies equally to a mouse cursor, a spoken command, and a pixel on a screen; and (2) a survey of the channels that recur in HCI, visualization, and ubiquitous analytics.
We give the first in \S\ref{sec:channel-properties} and the second in \S\ref{sec:output-channels} and \S\ref{sec:input-channels}.
Rather than an exhaustive enumeration---HCI has produced many, from Buxton's dimensionality taxonomy~\cite{Buxton1983lexical} through Mackinlay et al.'s input device design space~\cite{Card1990input}---our contribution is a uniform vocabulary: the same properties applied to input and output, to physical and digital channels, which is what makes the reanalysis in Section~\ref{sec:analysis} possible.

\begin{figure*}[htb]
    \centering
    \includegraphics[width=\linewidth]{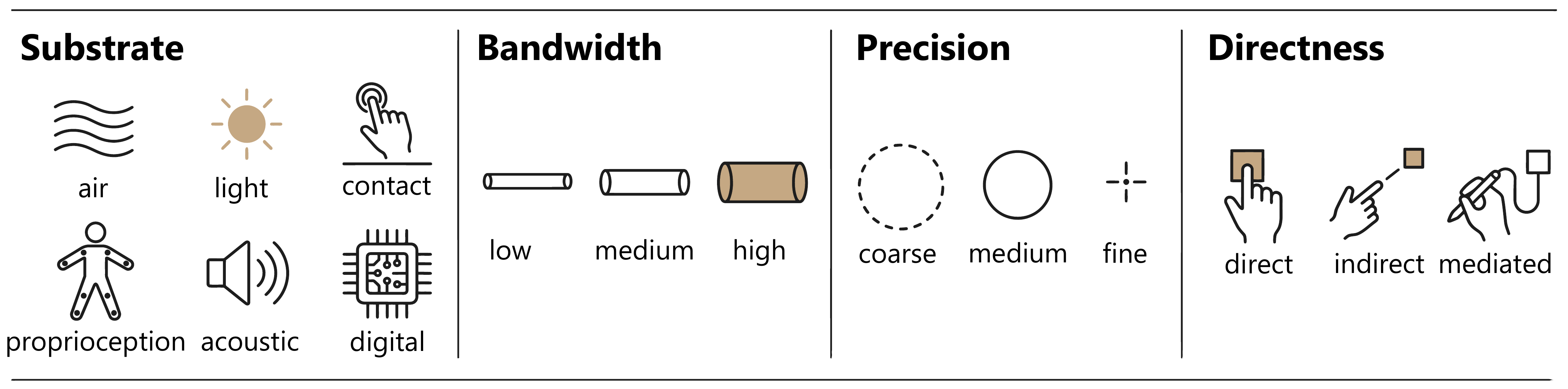}
    \caption{%
    \textbf{Input and output channel design space.}
        Each of the four analytical dimensions used to characterize input and output channels.
        \textit{Substrate} distinguishes six propagation media (this is a non-exhaustive list): air, light, contact/touch, proprioception, acoustic, and digital.
        \textit{Bandwidth} represents the rate at which representational state can pass through the channel.
        \textit{Precision} represents the granularity at which the channel can specify values or targets.
        \textit{Directness} represents the progression from direct action, through indirect action, to action mediated by a symbolic or instrumental layer.
    }
    \Description{A four-panel legend divided by vertical lines, with each panel headed by the name of one analytical dimension. The first panel, Substrate, contains six small icons arranged in two rows of three. Top row: three wavy horizontal lines labeled air; a sun with radiating rays labeled light; a fingertip touching a horizontal surface labeled contact. Bottom row: a simplified human figure with small dots at the joints labeled proprioception; a loudspeaker emitting curved sound waves labeled acoustic; a square chip with circuit traces labeled digital. The second panel, Bandwidth, contains three horizontal cylindrical pipes of increasing diameter from left to right, labeled low, medium, and high; the high-bandwidth pipe is filled with a tan color while the other two are outlined only. The third panel, Precision, contains three target shapes decreasing in size from left to right: a large dashed circle labeled coarse, a smaller solid circle labeled medium, and a small crosshair labeled fine. The fourth panel, Directness, contains three hand-and-target diagrams: a hand with the index finger pressing a tan square labeled direct; a hand pointing with the index finger at a small tan square separated by a short dashed line labeled indirect; a hand holding a stylus connected by a wire to a small outlined square labeled mediated.}
    \label{fig:design-space}
\end{figure*}

\subsection{Channel Properties}
\label{sec:channel-properties}

Figure~\ref{fig:design-space} organizes the vocabulary along four dimensions.
The first, \textit{substrate}, is a reference to its parent so that every channel is an attribute of some substrate, and the substrate determines most of what the channel can and cannot do.
We distinguish common substrates that recur in ubiquitous analytics: \textit{air} (free-space gestures sensed by cameras or radar), \textit{light} (graphical output to the eye and gaze input recovered from pupil geometry), \textit{contact} (touch, stylus, and keyboard on a shared surface), \textit{proprioception} (head and body pose, externalized through tracking), \textit{acoustic} (sound resolved by the ear, from speech to sonification), and \textit{digital} (pixels, memory, packets, sensor readings).
Large-scale interaction means bringing substrates into coordination with each other: a voice command crosses air, acoustic, and digital substrates in sequence, and each crossing is an interface at which representational state can be lost or transformed.

A single substrate generally hosts a family of channels rather than just one.
Visualization makes this obvious~\cite{Munzner2014}: the pixel grid is a single substrate, but position, length, area, color hue, color value, orientation, and motion are separate channels on it, with characteristic rank orderings for different data types~\cite{Cleveland1984, Mackinlay1986}.
The same holds elsewhere.
Air hosts gesture trajectory, gesture shape, and gesture velocity as distinct channels that can be read independently.
Speech hosts phonetic content, prosody, and timbre.
Channels within a substrate are often co-dependent---color hue cannot be varied without occupying position, and trajectory cannot be produced without also producing velocity---which is a source of both expressive power and interference.

The remaining three dimensions describe how information moves through a channel.
\textit{Bandwidth} is the rate at which representational state passes through, measured loosely as low, medium, or high.
A mouse delivers a medium-rate 2D stream bounded by hand motion; speech delivers a low- to medium-rate sequential stream capped by articulation at roughly 150 words per minute; gaze can fixate three or four targets per second and is high-rate; and vision, as output, parses millions of points in parallel per fixation, which is why visualization works at all~\cite{Card1999}.
\textit{Precision} is the granularity at which the channel can specify a value or target, and it is not a fixed property of the substrate: the contact substrate hosts a stylus tip (fine), a fingertip (medium), and a palm (coarse); the light substrate hosts a foveated fixation (medium) and a gross head orientation (coarse).
Precision is therefore a property of the channel-as-configured, which is one reason we do not identify channel with substrate.
\textit{Directness} follows Beaudouin-Lafon's degree of indirection~\cite{Beaudouin-Lafon2000}: a channel is \textit{direct} when action happens in the same spatial frame as the representation (a finger on a touchscreen, a hand on a virtual object); \textit{indirect} when an offset or scaling separates the two frames (a mouse moves a cursor on a separate display); and \textit{mediated} when a symbolic layer intervenes (speech names a target; a keyboard types letters a parser assembles).

\begin{table}[tb]
  \centering
  \small
  \caption{%
    \textbf{Summary of output and input channels.}
    Each channel listed with their parent substrate and intrinsic properties.    
    Output channels (\textcolor{channelOutput}{\faCircle}) carry information from the system to the analyst; input channels (\textcolor{channelInput}{\faCircle}) carry information from the analyst to the system.
    Entries give characteristic values; actual ranges depend on hardware and configuration.
  }
  \label{tab:channels}
  \begin{tabular}{@{}c l l l l l@{}}
    \toprule
    & \textbf{Channel} & \textbf{Substrate} & \textbf{Bandwidth} & \textbf{Precision} & \textbf{Directness} \\
    \midrule
    \multicolumn{6}{@{}l}{\textit{Output channels}} \\[2pt]
    \oc{visual}    & visual    & light / digital  & high         & fine        & direct/indirect \\
    \oc{audio}     & audio     & acoustic         & low          & coarse      & indirect \\
    \oc{haptic}    & haptic    & contact          & low          & coarse      & direct \\
    \oc{olfactory} & olfactory & air              & low          & coarse      & mediated \\
    \midrule
    \multicolumn{6}{@{}l}{\textit{Input channels}} \\[2pt]
    \ic{pointing}  & pointing  & contact          & medium       & fine        & direct/indirect \\
    \ic{touch}     & touch     & contact          & low--med     & medium      & direct \\
    \ic{voice}     & voice     & air / acoustic   & medium       & medium      & mediated \\
    \ic{gaze}      & gaze      & light            & high         & medium      & direct \\
    \ic{gesture}   & gesture   & air              & med--high    & medium      & direct/indirect \\
    \ic{body}      & body      & proprioception   & low          & coarse      & direct \\
    \ic{tangible}  & tangible  & contact          & low          & medium      & direct \\
    \ic{keyboard}  & keyboard  & contact          & high         & fine        & mediated \\
    \bottomrule
  \end{tabular}
\end{table}

\subsection{Output Channels}
\label{sec:output-channels}

Output channels carry information from the system to the analyst.
They are the substrate side of what visualization calls \textit{visual channels} when the output is graphical~\cite{Bertin1967, Card1999, Munzner2014}, and they generalize to the non-graphical case.
Our treatment here is brief: the visualization literature has covered the dominant output substrate in depth, and our interest is to reframe it as a special case of the same vocabulary we use for input.

\paragraph{\ocL{visual} output.}

The pixel grid is the dominant output substrate across data visualization and classical graphical interfaces, from desktop monitors and wall-sized displays to tablets, phones, watches, and head-mounted displays.
Its bandwidth is the highest of any channel we have, its precision is pixel-limited, and its directness depends on where the pixels live: a head-mounted display is direct with respect to gaze, while a wall display is direct with respect to group-level attention.
The substrate hosts many channels simultaneously---Munzner counts roughly a dozen on a 2D screen~\cite{Munzner2014}---and immersive settings add depth, binocular disparity, head-parallax motion, and spatial layout in the room.

\paragraph{\ocL{audio} output.}

Audio is a low-bandwidth, sequential, spatially vague output on the acoustic substrate, with virtues complementary to vision: it reaches the analyst who is not looking, survives occlusion and low light, and conveys urgency through timbre and rhythm.
Three uses recur.
Alarms and notifications carry a few bits at high urgency.
Spatialized audio in AR and VR locates sources in 3D and is direct with respect to directional hearing.
Sonification~\cite{Hermann2011} maps data to audio parameters (pitch, tempo, timbre), trading vision's parallelism for the temporal structure audio handles well.
The latter has been particularly useful as a sensory substitution~\cite{cook2014assistive, DBLP:journals/tvcg/ChunduryPRTLE22} for blind and low vision (BLV) individuals needing to access data-rich representations~\cite{DBLP:journals/cgf/ChoiJPCE19, DBLP:journals/interactions/Elmqvist23}.

\paragraph{\ocL{haptic} output.}

Haptic output carries force, vibration, or texture through the contact substrate, and sometimes through proprioception when a worn device resists or guides motion~\cite{MacLean2008}.
It has been suggested as a complementary modality for data representation, as well as an alternate one for BLV users~\cite{DBLP:journals/toh/PaneelsR10}.
A related approach, data physicalization, dispenses with the device entirely: the data is encoded in the shape and material of a tangible artifact~\cite{DBLP:conf/chi/JansenDIAKKSH15}.

Nevertheless, haptic or physical representations share the sense of touch, proprioception, and force.
Bandwidth is low and precision is coarse for most commodity hardware, but directness is high: a pulse on the wrist is unambiguously here, now, and about this action.
Haptics fills a specific gap---feedback when the analyst cannot look at the display or hear the speaker, or when a confirmation should not compete for visual attention---which matters on watches, in AR controllers, and in collaborative settings where extra visual output would be disruptive.

\paragraph{\ocL{olfactory} and gustatory output.}

Smell and taste have been explored as output channels for data~\cite{Patnaik2019, DBLP:conf/chi/BatchPAE20}, listed here for completeness.
They are high-latency, low-bandwidth, and hard to reset between samples, and their role in analytical tasks is narrow.
Again, its main use is as a complement or, potentially, as an alternate during sensory substitution~\cite{DBLP:journals/tvcg/ChunduryPRTLE22}.

\subsection{Input Channels}
\label{sec:input-channels}

Input channels carry information from the analyst to the system.
Unlike the output side, where the pixel grid dominates, the input side is genuinely plural: different substrates fit different tasks and different settings, and no single channel plays the default role the screen plays for output.
We walk through the channels that recur in the UA literature and that the systems in Section~\ref{sec:analysis} use.

\paragraph{\icL{pointing} (mouse, trackpad, stylus, HMD controllers).}

Pointing is the classical input channel: a 2D/3D position stream on a flat or volumetric substrate: high precision, medium bandwidth, indirect or mediated depending on hardware~\cite{Shneiderman2016dtui, DBLP:journals/tois/CardMR91}.
A mouse on a desk is indirect; a stylus on a tablet is direct in position and indirect in tilt.
Pointing is the channel on which the WIMP interface~\cite{vanDam1997} was built and remains the most precise commodity input we have.
In ubiquitous settings it persists on tablets and laptops and fades on phones, watches, and headsets, where the physical surface is not available in the same way.

\paragraph{\icL{touch}.}

Touch is a 2D position stream on the contact substrate, medium precision, low to medium bandwidth, direct by construction~\cite{Potter1988}.
Multi-touch~\cite{Lee1985, Wu2003} expands the bandwidth through parallel fingers---pinch, rotate, two-handed gestures---or extended above the touch surface~\cite{DBLP:conf/uist/MurugappanVER12}.
However, the ten-finger ceiling caps this sharply: past three or four fingers, recognition and recall both collapse.
The directness of touch makes it excellent for selection and manipulation of on-screen objects but poor for text entry~\cite{DBLP:conf/chi/VogelB07} (where the finger occludes the very keys it presses) and poor for anything off the surface.

\paragraph{\icL{voice}.}

Voice is a sequential acoustic channel: medium bandwidth bounded by articulation rate, and medium to high precision for in-vocabulary utterances.
It is mediated because it refers to targets rather than pointing at them.
Voice is the input channel that most closes the goal-formation gulf~\cite{lam2008}: an analyst can say \textit{``show me stores where returns are up,''} and a competent system can enact that intent without the analyst mapping it to a specific widget~\cite{DBLP:conf/uist/SrinivasanS21, DBLP:journals/tvcg/SrinivasanS18}.
It suffers in noise, suffers socially in quiet or shared spaces, and fails on deictic or spatial references (\textit{``that one, no, the one next to it''}) that gesture or gaze handles trivially.

\paragraph{\icL{gaze}.}

Gaze is a high-rate pointing stream on the light substrate, medium precision, high bandwidth, and direct with respect to the visual scene.
Gaze serves double duty: it is both an implicit indicator of visual attention and an explicit input mechanism, and researchers have exploited both since Bolt's gaze-orchestrated windows~\cite{Bolt1980} (which we will discuss below).
Gaze is cheap: the analyst tends to look at what they are thinking about anyway, so gaze-as-input piggybacks on attention with no extra motor cost.
It also has the Midas-touch problem~\cite{Jacob1991}: if every fixation is an action, the analyst cannot just look at things without triggering them, and disambiguating fixations from commands requires either dwell time (slow) or a second channel for confirmation~\cite{DBLP:conf/ngca/StellmachSND11, DBLP:conf/uist/PfeufferACG14}.
In current headsets, gaze-plus-pinch has become the default selection channel on the Apple Vision Pro and Android XR, which illustrates our complementarity point: gaze provides the pointer; pinch provides the commit.

\paragraph{\icL{gesture}.}

Gesture is a free-space input channel on the air substrate, tracked by cameras, radar, or worn sensors~\cite{Bolt1980}.
Bandwidth is medium to high for expressive two-handed gestures; precision is medium at close range and degrades with distance; directness depends on whether the gesture happens at the target or in a body-relative frame that maps onto it.
The Midas-touch problem returns here as \textit{gorilla arm}~\cite{HincapieRamos2014}: sustained mid-air gesture fatigues the deltoid within minutes, which makes gesture excellent for brief expressive acts and poor for continuous control.
We argue that fatigue, not precision, is the dominant design constraint on gesture-as-input, and that shallow-depth interaction~\cite{Hancock2007} and body-relative framing are the two moves that actually work.

\paragraph{\icL{body} position and locomotion.}

Head pose, body orientation, and foot position are input channels on the proprioception substrate, externalized through tracking.
Bandwidth is low but directness is very high: the analyst's location in the room \textit{is} the query~\cite{Ball2007b}.
Walking around a 3D scatterplot inspects different regions; stepping toward a display signals interest; turning away ends interaction.
These are the channels proxemic interaction work has pursued~\cite{Greenberg2011}, and the ones ubiquitous analytics turns on most~\cite{Jakobsen2013, Badam2016b}.
Body-as-input is free---the analyst was going to move anyway---but only if the system interprets the movement in the analyst's frame rather than the system's.

\paragraph{\icL{tangible} and physical tools.}

Tangibles are input channels with a physical artifact as part of the substrate: a block on a tabletop, a prop in the hand, a paper card on a sensing surface~\cite{DBLP:conf/chi/IshiiU97, Shaer2010}.
Bandwidth is low and precision is medium, but directness is excellent because the object is the referent.
In analytics, tangibles have served as filters, views, axes, and role tokens.
These tangibles can also serve as the input channel equivalent of data physicalization~\cite{DBLP:conf/chi/JansenDIAKKSH15} if instrumented, such as for stacking physical tokens~\cite{DBLP:conf/avi/KlumILFD12}, manipulating them as movable views over a tabletop dataset~\cite{Spindler2010, Kim2012}, or actuating them as swarms that both display and respond to data~\cite{LeGoc2016}.

\paragraph{\icL{keyboard} and text.}

The keyboard is a high-bandwidth, high-precision, mediated input channel on the contact substrate, and for text entry it has no rival in any current setting~\cite{DBLP:journals/tois/CardMR91, Shneiderman2016dtui}.
In ubiquitous analytics the keyboard persists at the desktop and struggles everywhere else, exactly the availability-versus-suitability distinction developed in Section~\ref{sec:model}.

\section{Analysis: System Examples}
\label{sec:analysis}

We now apply the channel/substrate framework to reanalyze several ubiquitous, immersive, and situated analytics systems.
We chose these systems as representative examples of immersive, situated, and ubiquitous analytics that employ distinct combinations of input and output channels.
\autoref{tab:modalities} summarizes the systems and their channel support.

\begin{table*}[htb]
    \centering
    \small
    \caption{\textbf{Mapping of output and input channels.}
    Shown across a sampling of ubiquitous, immersive, and situated analytics systems.
    $\checkmark$ denotes a primary modality, $\circ$ a secondary or optional modality, and -- a modality not supported or not described.}
    \label{tab:modalities}
    \begin{tabular}{lcccccccccccc}
        \toprule
        \textbf{System} 
        & \multicolumn{4}{c}{\textbf{Output}} 
        & \multicolumn{8}{c}{\textbf{Input}} \\
        \cmidrule(lr){2-5} \cmidrule(lr){6-13}
        & \oc{visual} & \oc{audio} & \oc{haptic} & \oc{olfactory}
        & \ic{pointing} & \ic{touch} & \ic{voice} & \ic{gaze} & \ic{gesture} & \ic{body} & \ic{tangible} & \ic{keyboard} \\
        \midrule

        Put-That-There~\cite{Bolt1980}  
        & $\checkmark$ & $\circ$ & -- & -- 
        & $\checkmark$ & -- & $\checkmark$ & -- & -- & -- & -- & -- \\
                
        ImAxes~\cite{cordeil17imaxes}  
        & $\checkmark$ & -- & -- & -- 
        & $\checkmark$ & -- & -- & -- & -- & -- & -- & -- \\
        
        DashSpace~\cite{Borowski2025dashspace}  
        & $\checkmark$ & -- & -- & -- 
        & $\checkmark$ & $\circ$ & $\circ$ & $\checkmark$  & $\circ$ & $\checkmark$ & -- & $\circ$ \\
        
        Wizualization~\cite{DBLP:journals/tvcg/BatchBRE24}
        & $\checkmark$ & -- & -- & -- 
        & -- & -- & $\checkmark$ & $\checkmark$ & $\checkmark$ & -- & -- & -- \\
        
        
        MARVIS~\cite{langner21marvis}       
        & $\checkmark$ & -- & -- & -- 
        & $\checkmark$ & $\checkmark$ & -- & $\checkmark$  & -- & $\circ$ & $\circ$ & $\checkmark$ \\
        
        
        
        NoSpoon~\cite{batch20econimmersive}     
        & $\checkmark$ & -- & -- & $\checkmark$
        & $\checkmark$ & -- & -- & $\checkmark$  & -- & -- & -- & -- \\
        
        Munin~\cite{Badam2015}      
        & $\checkmark$ & -- & -- & -- 
        & $\checkmark$ & $\checkmark$ & -- & -- & -- & -- & -- & $\checkmark$ \\
        \bottomrule
    \end{tabular}
\end{table*}

\begin{figure*}[htb]
    \centering
    \includegraphics[width=\linewidth]{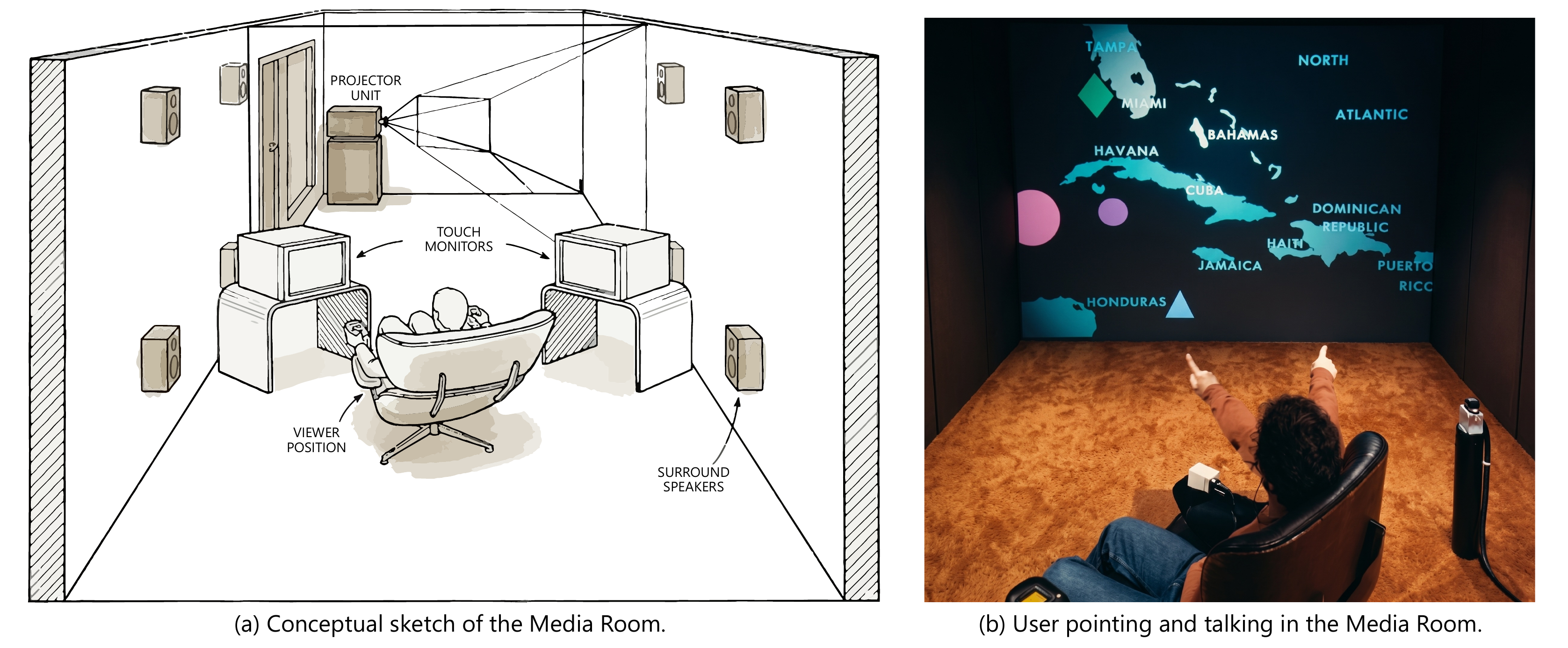}
    \caption{\textbf{Put That There~\cite{Bolt1980}.}
    The MIT Media Room and its SDMS combined gestural and voice interaction system. 
    The left image (a) shows the conceptual sketch, the right (b) shows the system in actual use.
    }
    \label{fig:put-that-there}
    \Description{Put-That-There system in the MIT Media Room. Two panels: (a) a conceptual sketch showing a room-sized installation with a rear-projection screen on the far wall, a seated user in an Eames-style chair at center, touch monitors flanking the chair, surround speakers mounted on side and rear walls, and a ceiling-mounted projector unit; (b) a photograph of the system in use, showing a seated user pointing with both hands toward a wall-sized display showing colored circles, diamonds, and triangles overlaid on a Caribbean map, with a Polhemus magnetic transmitter cube visible on a pedestal to the right of the chair.}
\end{figure*}

\subsection{Put-That-There}

Bolt's ``Put-That-There''~\cite{Bolt1980} (officially known as the Spatial Data-Management System, SDMS) is among the earliest demonstrations of multimodal input in a spatial computing environment (Figure~\ref{fig:put-that-there}).
Built in MIT's Media Room---a room-sized terminal with a wall-sized rear-projected display---the system let a seated user create, move, copy, and delete colored shapes on the large screen by speaking commands while pointing with one hand.
A wrist-mounted Polhemus magnetic sensor tracked the hand's 3D position in the room; a connected-speech recognizer (NEC DP-100, 120-word vocabulary) parsed short sentences of up to five words.
The two input channels served complementary roles.
Voice specified the \emph{action} and the \emph{attributes}: ``create a blue square,'' ``make that smaller.''
Pointing specified \emph{where} and \emph{which}: the spatial referent that voice alone could not resolve.
Pronouns---``that,'' ``there''---bridged the two channels, functioning as deictic anchors that bound the voice stream to the pointing stream at the moment of utterance.

\paragraph{Output channels.}

Put-That-There uses a single output channel: 2D \ocL{visual}.\footnote{Note that, as Figure~\ref{fig:put-that-there} shows, the Media Room does come equipped with six surround speakers, and the environment is described as ``arich graphics world of color and sound.'' 
Nevertheless, Bolt's paper~\cite{Bolt1980} does \textbf{not} explicitly discuss the use of \ocL{audio} output in the SDMS system.}
The wall-sized rear-projection screen displays the shapes, a small white ``x'' cursor tracks pointing position, and addressed items briefly desaturate to confirm selection.
When the user says ``put that there'' while pointing, the object moves to the pointed-at location, and the visual feedback is immediate and co-located with the action.

\paragraph{Input channels.}

The system uses two input channels in concert: \icL{pointing} and \icL{voice}.\footnote{Again, the Media Room has joysticks and touch displays, but SDMS does not seem to utilize them.}
The pointing channel operates via a wrist-mounted magnetic position sensor that reports 3D coordinates in the room; a ray from the sensor to the display surface maps hand position to screen coordinates.
The voice channel operates via connected-speech recognition, constrained to short imperative sentences.
Pointing has high spatial directness---the hand position maps onto a screen location with minimal translation---but low symbolic bandwidth: it can indicate \emph{where}, not \emph{what to do}.
Voice has high symbolic bandwidth---it specifies actions, attributes, and object types---but near-zero spatial precision: saying ``to the right of the green square'' is imprecise and presupposes a reference object that may not exist.

\begin{figure*}[htb]
    \centering
    \includegraphics[width=\linewidth]{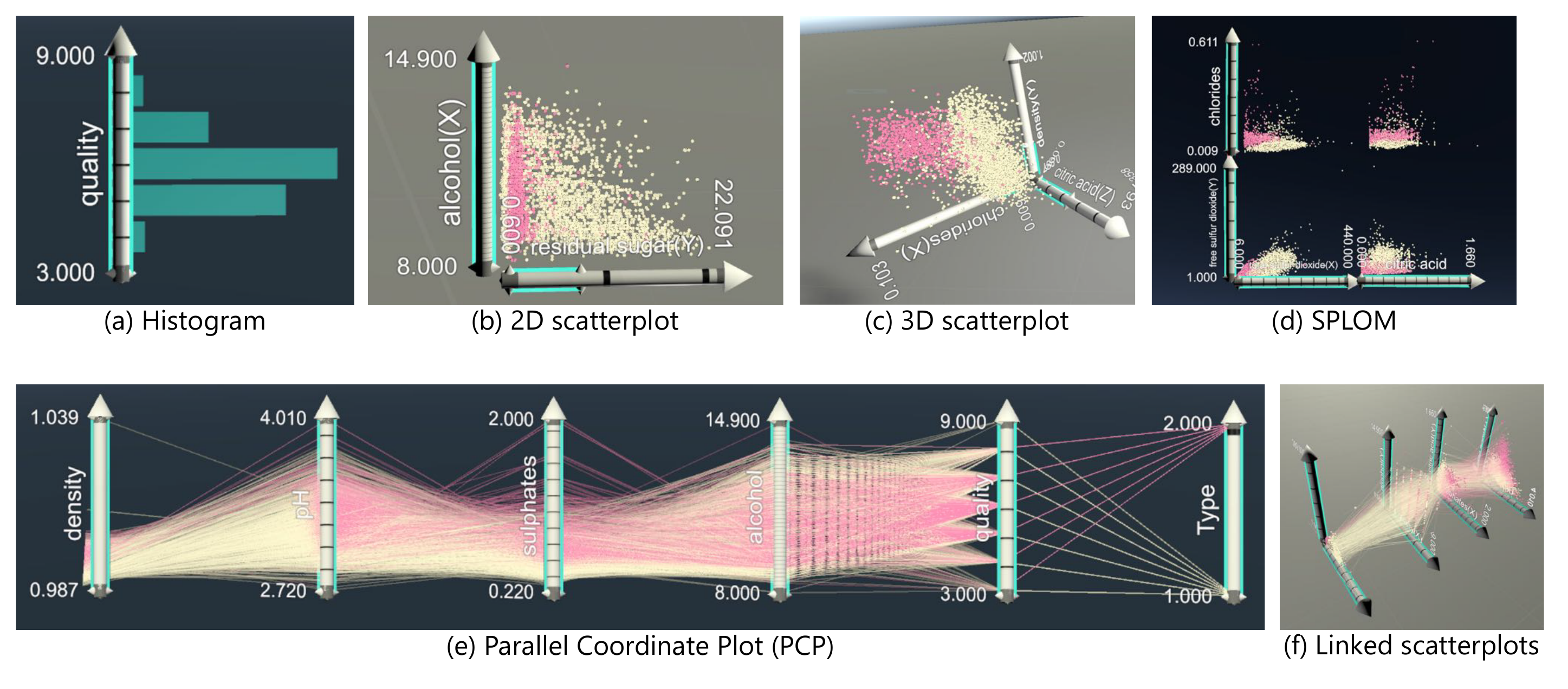}
    \caption{\textbf{ImAxes~\cite{cordeil17imaxes}.}
    One of the first immersive analytics system, ImAxes supports multidimensional visual exploration using an intuitive 3D manipulation interaction accessed using 3D motion controllers in immersive Virtual Reality.
    }
    \label{fig:imaxes}
    \Description{ImAxes immersive analytics system showing six coordinated visualization types in Virtual Reality. Six panels labelled (a) through (f) show different chart types rendered as 3D objects manipulated via motion controllers. (a) Histogram of quality with horizontal bars. (b) 2D scatterplot of alcohol vs. residual sugar. (c) 3D scatterplot adding chlorides as a third axis. (d) Scatterplot matrix (SPLOM) arranged as a 3D grid of small multiples. (e) Parallel coordinate plot with five axes (density, pH, sulphates, alcohol, quality, type) showing dense crossings between axes. (f) Linked scatterplots composed by physically arranging multiple axes in 3D space, with brushed selections highlighted in pink across views. All visualizations use a dark background with data points in yellow-green and selections in pink.}
\end{figure*}

\subsection{ImAxes}
\label{sec:imaxes}

ImAxes~\cite{cordeil17imaxes} is a VR system for multivariate data exploration in which each data dimension is represented as a 3D axis object that the analyst can grab, reposition, and combine (Figure~\ref{fig:imaxes}).
The type of visualization that appears---scatterplot, parallel coordinates, histogram, or linked hybrid---is determined entirely by the spatial arrangement of axes relative to one another; a formal grammar maps axis proximity and orientation to visualization type.
No menus or mode switches are required.
This makes ImAxes a clean case of spatial directness in the channel/substrate framework: the analyst's hand position in the virtual environment maps directly onto the representation space, and the axes themselves function as external representations in the DCog sense; persistent, manipulable artifacts that offload the combinatorial work of exploring dimension pairings from working memory onto the spatial environment.

\paragraph{Output channels.}

ImAxes relies on a single output channel: 3D \ocL{visual}.
The VR headset renders axes, data marks, and linking geometry in stereo 3D with head tracking, and the analyst reads the resulting visualizations by looking at and walking around them.
The controllers do provide a brief haptic buzz when an axis is pulled past the duplication threshold on the attribute shelf, but this is feedback on the input action, not a channel for data output.
This single-channel output profile is typical of immersive analytics systems: the bandwidth and precision of stereo vision in a head-tracked environment are high enough that one output channel suffices for the analytical task.

\paragraph{Input channels.}

ImAxes uses a single input channel: 3D \icL{pointing} via 6DOF tracked VR controllers (HTC Vive).
The analyst grabs an axis by pressing the controller trigger while it intersects the axis, repositions it, and releases.
In the terminology of our framework, the pointing channel here has high directness---the controller's 6DOF position maps one-to-one onto the axis position in the data space---and high bandwidth, since the analyst simultaneously specifies position and orientation.
Precision is adequate for the coarse spatial grammar (axis proximity and angle thresholds), though it would not suffice for fine-grained value selection or text entry.

The channel profile of ImAxes is narrow by design: one input channel, one output channel.
The system's contribution to this analysis is the directness of the single channel pair it employs.
The directness is high and the execution gulf is small because grabbing and positioning a virtual object is a spatial action that maps onto the analyst's intention with minimal translation.
The evaluation gulf is similarly narrow: the visualization appears immediately at the axes' location, so the analyst evaluates the result by looking at the same spatial region where the action was performed.
The emergent interactions that Cordeil et al.\ report---axis swiping, brushing with motion, embodied queries where a filtered visualization becomes a tangible brush---all derive from this directness: because the channel maps action onto representation space without indirection, composing actions composes representations.

\begin{figure*}[th]
    \centering
    \includegraphics[width=\linewidth]{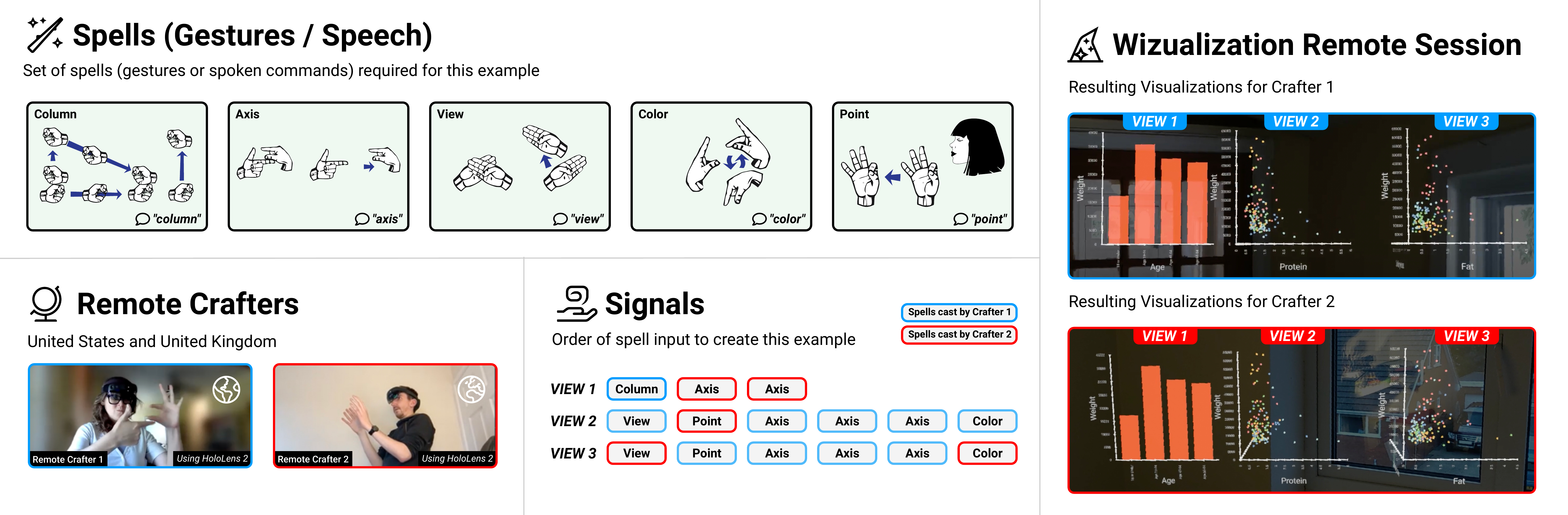}
    \caption{\textbf{Wizualization~\cite{DBLP:journals/tvcg/BatchBRE24}}.
    The system uses a combination of \icL{gesture} (using ASL), 3D \icL{pointing}, and \icL{voice} commands to enable remote users to co-create visualizations and co-analyze them in AR space.}
    \Description{Wizualization system overview in four panels. Top left: six ASL-derived hand gestures called ``spells'' used as input commands, labelled Column, Axis, View, Color, and Point, each shown as a hand illustration paired with its spoken equivalent. Bottom left: two remote collaborators (``crafters'') in two different locations, each wearing a HoloLens 2 headset. Bottom center: a signal sequence showing the order of spells cast by each crafter to co-create three coordinated views, with colored labels distinguishing contributions from Crafter 1 and Crafter 2. Right: the resulting AR visualizations as seen by each crafter, showing three linked views including bar charts and a scatterplot of a nutrition dataset with axes for Protein and Fat.}
    \label{fig:wizualization}
\end{figure*}

\subsection{Wizualization}

Wizualization~\cite{DBLP:journals/tvcg/BatchBRE24} is an immersive and ubiquitous analytics system that uses a ``hard magic'' metaphor (i.e., following strict rules) for authoring of 3D visualizations in immersive space (Figure~\ref{fig:wizualization}).
The system explores how visualizations can be created  through consistent, learnable, and rule‑based gestures and voice commands.
It integrates AR, gestural interaction using an ASL-based language, and a grammar of graphics for immersive visualizations and interactions, for creating the former.
Wizualization supports collaborative data analysis across physical and digital spaces, and, rather than focusing on visual encoding innovations, emphasizes the orchestration of interaction techniques and perceptual channels.

\paragraph{Output channels.} 

Wizualization uses a single output channel: 3D \ocL{visual}.
Using AR displays, visualizations are persistently situated in space. Visual output is enhanced with by dynamic animations that communicate interactions and events.
While vision dominates, output is not purely graphical: spatial arrangement, scale, and physical alignment act as cues that convey semantic structure, which along with the dynamic animations, are very important when building visualizations collaboratively.
Similarly to ImAxes, and most IA/SA systems, Wizualization uses a single-channel output. 

\paragraph{Input channels.} 

User input in Wizualization is via 3D \icL{pointing}, \icL{voice}, and \icL{gesture} commands.
Mid‑air hand gestures support direct manipulation of visualizations their elements (e.g., axis) and other attributes (e.g., colors), enabling operations such as selection, filtering, grouping, and transformations.
ASL gestures are also mapped to voice commands.
In addition to explicit interactions, the system leverages implicit input derived from user position, orientation, and proximity, allowing analytical context to evolve as users move through collaborative space. 

Unlike ImAxes, the channel profile is a bit broader as there are two input modalities that are being used.
Moreover, due to the AR nature of the output, and the collaborative support, the co-presence of users and the graphical depiction of their actions from each collaborators point of view are important elements of the system's functionality.
The execution gulf is somewhat larger for this reason.
Consequently, so is the evaluation gulf, as the creation of visualization from one participant takes some time to be completed.
The other participants may or may not be fully aware of the visualization creators intention as the visualization is crafted. 

\begin{figure*}[htb]
    \centering
    \includegraphics[width=\linewidth]{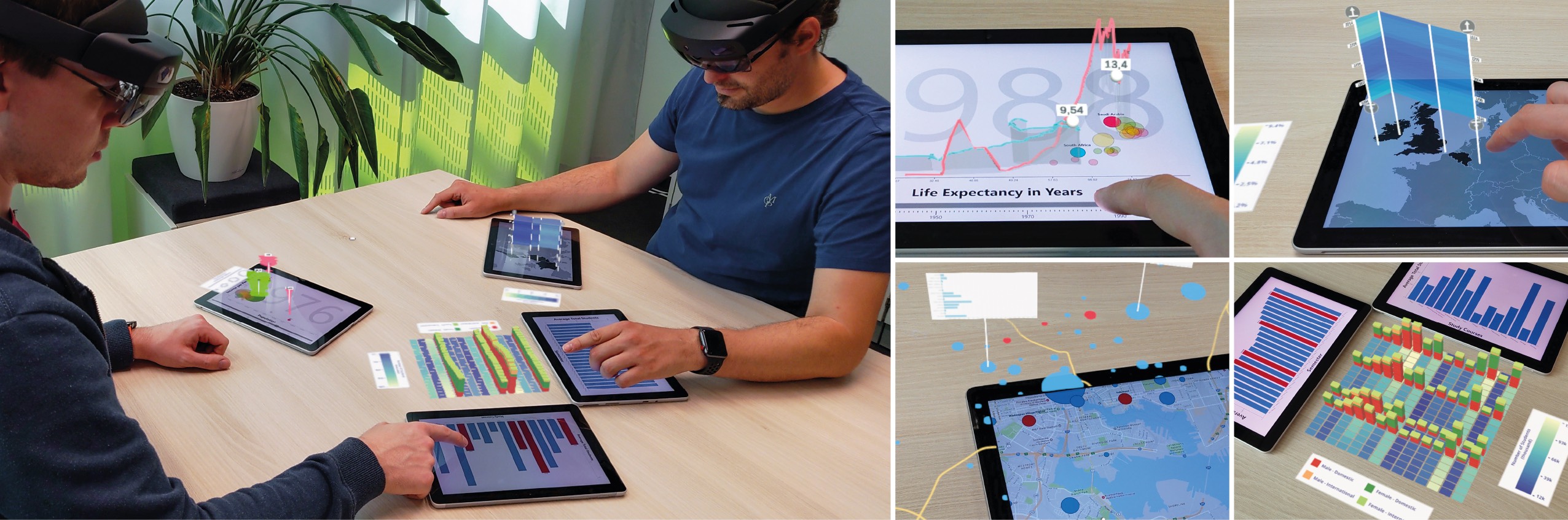}
    \caption{\textbf{MARVIS~\cite{langner21marvis}}.
    This system combines mobile devices with AR HMDs to create a collaborative visualization space.
    Visualizations are displayed via the AR HMD adjacent or superimposed to the mobile devices to augment the display capabilities of the latter.
    }
    \label{fig:marvis}
    \Description{MARVIS system combining tablets and AR headsets. Left: two users wearing HoloLens headsets sit at a table with multiple iPads showing bar charts and parallel coordinates; AR-rendered 3D visualizations float above the tablets. Top center: close-up of a tablet displaying a life expectancy bar chart with AR annotations and linking lines extending above the screen. Top right: a user's hand tilts a tablet showing a world map, with AR content attached to the device surface. Bottom left: a tabletop arrangement of tablets showing a geographic map with bar charts distributed across multiple devices, connected by AR linking lines. Bottom right: two tablets side by side displaying treemaps and bar charts with shared color encoding.}
\end{figure*}

\subsection{MARVIS}

MARVIS~\cite{langner21marvis} combines AR headsets with handheld devices to support visual data analysis (Figure~\ref{fig:marvis}).
The tool integrates familiar touch‑based interaction with spatially situated visualization, allowing users to fluidly transition between detailed, private views and large‑scale, shared representations.
The primary feature that makes MARVIS unique is the extension of the viewing area of, say, a tablet by superimposing coordinated 3D visualizations on the devices, or 2D and 3D visualizations adjacent to said tablet.

\paragraph{Output channels.} 

MARVIS uses \ocL{visual} output is as the primary output, split between private, high‑resolution displays on mobile devices and shared, spatially embedded visualizations presented through AR.
The mobile device functions as a personal output channel, supporting detailed inspection, annotation, and focused analysis.
AR, by contrast, provides a large‑scale spatial output channel in which visualizations can be externalized, spatially arranged, and collaboratively explored.
Visual continuity between these channels is maintained through coordinated views and synchronized state, ensuring that analytical context is preserved across devices.
As in ImAxes, spatial layout and viewpoint play an important role in conveying structure, with physical positioning of the handhelds acting as an implicit visual variable.

\paragraph{Input channels.} 

MARVIS distributes input across both \icL{touch} and \icL{keyboard} through the handheld devices, which serves as the primary explicit input mechanism, supporting precise selection, filtering, and parameter manipulation using gestures.
In parallel, \icL{tangible} and spatial input is provided through user movement and device placement within the AR environment, enabling users to reposition views, inspect data from different perspectives, and interact with shared visualizations through physical navigation. 

The channel profile for MARVIS is more complex, due to the combination of different device types affecting input and output.
The \icL{visual} output remains the dominant modality, but is partitioned across personal and shared displays to support different analytical needs. 
The execution gulf is wider, with input distributed, across devices and collaborators, yet supported though the tight synergy between handhelds and AR headsets.
As each device type complements each other, a core aspect of the sensemaking process how each device type complements each other in the interactive and visual space.
On the other hand this synergy also influences the gulf of evaluation, as the sensemaking process is facilitated by the shared exploration of the augmented space.
Another important aspect is that the shared view in AR is different from the private view on collaborating handhelds, which widens both the gulfs of execution and evaluation.

\subsection{Summary: Ubiquitous, Immersive, and Situated Analytics Systems}

The remaining systems in \autoref{tab:modalities} reinforce the pattern of visual dominance on the output side while illustrating distinct input channel profiles.
DashSpace~\cite{Borowski2025dashspace} has the broadest input profile in our sample: \icL{pointing} and \icL{gaze} serve as primary channels, supplemented by optional \icL{touch}, \icL{voice}, \icL{gesture}, and \icL{keyboard}, with \icL{body} position as a primary spatial input, a breadth that reflects its design as a general-purpose MR analytics environment.
NoSpoon~\cite{batch20econimmersive} is the only system in our sample to use a non-visual output channel; beyond the visual one, it also supports \ocL{olfactory} output~\cite{Patnaik2019}.
Its input side is narrow---\icL{pointing} and \icL{gaze}---matching the exploratory rather than authoring focus of the system.
Munin~\cite{Badam2015} takes a cross-device approach similar to MARVIS but without the AR layer: \icL{touch} and \icL{keyboard} on personal devices combine with \icL{pointing} on a shared display, distributing input across device boundaries while keeping output purely \ocL{visual}.

\section{Discussion}
\label{sec:discussion}

This paper has taken the stance that distributed cognition can provide a useful framing for human-computer interaction, modeling all exchange between users and computational devices as information propagation through different \textit{interaction substrates}~\cite{Mackay2025substrates}.
Another useful piece of the puzzle can be found in the field of data visualization, where the notion of a \textit{visual channel} as an information-carrying attribute of the visual media can be generalized as \textit{input channels} and \textit{output channels} in a general HCI interaction model.
We have used this reenvisioned model to enumerate various input and output channels and then to reanalyze existing ubiquitous, immersive, and situated analytics systems.

Here we use these ideas as a springboard for discussing several extensions.
First of all, we discuss how they might influence future designs.
We then speculate about how distributed cognition can be also become a useful framing for discussing the integration of human-centered AI in these tools.
We close with a discussion of our limitations and future work. 

\subsection{Implications for Design}

Our channel and substrate model gives designers a vocabulary for reasoning precisely about what information is being conveyed, in what direction, and through which substrate.
This is useful in any HCI setting, but it is particularly useful for ubiquitous analytics, where the same analytical task may unfold across very different physical and social contexts.
Rather than asking ``which modality should we support?'' the model asks, which channels are \textit{available} in this setting, which are \textit{suitable} for the task at hand, and which are \textit{preferable} given the analyst's constraints?

More broadly, the DCog framing reminds us that interaction does not live inside a device but is distributed across the substrates that a setting affords.
This way of thinking is especially apt as computing devices get smaller, more numerous, and increasingly invisible.
We already interact with information this way outside of computing: we spread tax documents across the kitchen table, pin travel plans to the wall, read the newspaper next to our coffee.
These are spatial, multi-substrate information arrangements, and we navigate them without thinking about ``interaction techniques.''
The channel/substrate model makes explicit what these arrangements \textit{do}: they distribute representational state across physical substrates and let us propagate information between them using whatever channels are at hand---gaze, touch, spatial memory, speech.
Ubiquitous analytics should aim for the same fluency.

This view also suggests that the most productive design strategy is not to optimize individual channels in isolation, but to compose channel ensembles that complement each other's weaknesses.
Gaze is fast but imprecise; gesture is spatially expressive but fatiguing; voice is hands-free but spatially ambiguous.
A system that lets an analyst confirm a gaze-selected object with a pinch gesture, or disambiguate a voice command with a head turn, is exploiting channel complementarity, and the substrate model makes the rationale legible.

\subsection{Spatially Distributed Cognition for Human-Centered AI}

The DCog framing has an unexpected resonance with current discussions about AI in interactive systems.
The idea that intelligence can be distributed---embedded in the situation, the artifacts, the social arrangements, not concentrated in a single agent---is a DCog idea that predates modern AI by decades.
The intelligence in Hutchins's navigation was in the system, not in the individuals; in the context, not the plans~\cite{Suchman1987}; in space, not the mind~\cite{DBLP:journals/ai/Kirsh95}.
These are useful correctives to the prevailing AI integration paradigm, which tends to treat the AI as autonomous and independent agents inserted into the workflow.
A DCog perspective suggests a different approach: distribute agency into the environment through instruments and representations, and let intelligence emerge from the interactions between human participants, computational tools, and environmental context.

There is a suggestive parallel with how large language models and foundation models actually work.
Context is everything for these systems: the prompt, the surrounding conversation, the documents provided.
Context is everything for people too, but for people, context is overwhelmingly physical; the room they are in, the artifacts on the table, the colleague standing next to them.
Embedding computational context-awareness in the physical environment will essentially allow us to operationalize DCog through spatial computing and generative AI: computational intelligence that is not a separate agent to converse with, but a distributed property of the substrates we already inhabit.
It is an alterantive to a more situated and human-centered design.

\subsection{Limitations and Future Work}

The model we present is descriptive, not operational.
In this paper, we have used it as an analytical tool---to reanalyze existing systems, to identify what channels they exploit and what substrates they distribute cognition across---but we have not yet demonstrate its use as a generative tool.
That is left for future work.

Across the systems we examined, visual channels carry the bulk of the information from system to analyst; this is also true for the systems summarized in Table~\ref{tab:modalities}.
Audio, haptic, and proprioceptive output channels remain underexplored, even in systems designed for settings where visual attention is scarce or contested.
This is partly a reflection of the field's origins in data \textit{visualization}, but it is also a opportunity for future work in this field.
Settings like underwater fieldwork, factory floors with loud machinery, or emergency response under time pressure may benefit from output channel ensembles that do not privilege vision above all else.

\section{Conclusion}
\label{sec:conclusion}

In this paper we have argued that distributed cognition provides a productive framing for interaction in ubiquitous analytics, and that the visualization community's concept of a visual channel generalizes naturally into input and output channels defined as attributes of interaction substrates.
The resulting model offers a shared vocabulary---substrate, channel, bandwidth, precision, directness, and the available/suitable/preferable distinction---for reasoning about why certain interactions work in certain settings and fail in others.
Our reanalysis of existing systems demonstrates the model's descriptive power; the next step is to test its generative power by using it to drive new designs.

\begin{acks}
    This work was supported partly by Villum Investigator grant VL-54492 by Villum Fonden.
    Any opinions, findings, and conclusions expressed in this material are those of the authors and do not necessarily reflect the views of the funding agency.
\end{acks}

\section*{Use of Generative AI}

Google Gemini 3.1 Image Flash (Nano Banana 2) was used as a starting point for several figures in this paper. 
All figures have subsequently been manually revised and verified for accuracy.

\bibliographystyle{ACM-Reference-Format}
\bibliography{channels}

\end{document}